\documentclass[onecolumn,french,english]{article}
\usepackage[LGR,T1]{fontenc}
\usepackage[utf8]{inputenc}
\usepackage{geometry}
\usepackage{color}

\usepackage[left,modulo]{lineno}
\usepackage{setspace}

\usepackage{textcomp}
\usepackage{amstext}
\usepackage{graphicx}
\usepackage{esint}
\usepackage[authoryear]{natbib}
\usepackage{multicol}
\usepackage{upgreek}
\usepackage{abbrv}

\usepackage[]{hyperref}
\hypersetup{colorlinks=true,citecolor=blue}

\makeatletter

\DeclareFontEncoding{LGR}{}{}
\DeclareTextSymbol{\~}{LGR}{126}
\providecommand{\tabularnewline}{\\}

\makeatother

\usepackage{babel}
\makeatletter
\addto\extrasfrench{%
   \providecommand{\fg}{\ifdim\lastskip>\z@\unskip\fi~\frqq}%
}

\makeatother

\begin{document}

\title{Modeling the albedo of Earth-like magma ocean planets with H$_2$O-CO$_2$ atmospheres}

\author{$\mathrm{W.\,Pluriel}^{\text{1,2}}$, $\mathrm{E.\,Marcq}^{\text{2}}$, $\mathrm{M.\,Turbet}^{\text{3}}$}

\maketitle

$^\text{1}$ Laboratoire d'astrophysique de Bordeaux, Univ. Bordeaux, CNRS, B18N, allée Geoffroy Saint-Hilaire, 33615 Pessac, France
\\
$^\text{2}$ LATMOS/IPSL/CNRS/UPMC/UVSQ, Guyancourt, France
\\
$^\text{3}$ Laboratoire de Météorologie Dynamique, IPSL, Sorbonne Universités, UPMC Univ Paris 06, CNRS, 4 place Jussieu, 75005 Paris, France.


\begin{center}
\section*{Abstract}
\end{center}
During accretion, the young rocky planets are so hot that they become endowed with a magma ocean. From that moment, the mantle convective thermal flux control the cooling of the planet and an atmosphere is created by outgassing. This atmosphere will then play a key role during this cooling phase. Studying this cooling phase in details is a necessary step to explain the great diversity of the observed telluric planets and especially to assess the presence of surface liquid water.
We used here a radiative-convective 1D atmospheric model (H$_2$O, CO$_2$) to study the impact of the Bond albedo on the evolution of magma ocean planets. We derived from this model the thermal emission spectrum and the spectral reflectance of these planets, from which we calculated their Bond albedos. Compared to \citet{Marcqetal2017}, the model now includes a new module to compute the Rayleigh scattering, and state of the art CO$_2$ and H$_2$O gaseous opacities data in the visible and infrared spectral ranges.
We show that the Bond albedo of these planets depends on their surface temperature and results from a competition between Rayleigh scattering from the gases and Mie scattering from the clouds. The colder the surface temperature is, the thicker the clouds are, and the higher the Bond albedo is. We also evidence that the relative abundances of CO$_2$ and H$_2$O in the atmosphere strongly impact the Bond albedo. The Bond albedo is higher for atmospheres dominated by the CO$_2$, better Rayleigh scatterer than H$_2$O. Finally, we provide the community with an empirical formula for the Bond albedo that could be useful for future studies of magma ocean planets.

\textbf{Keywords:} Atmospheres; albedo; spectroscopy; radiative transfer; modeling 
\\
\section{Introduction }

In order to better assess the history of the atmospheres of telluric
planets, we need to better characterize the primitive atmospheres that
these planets possessed at the very beginning of their history. These
secondary atmospheres are outgassed during a relatively short stage
lasting $10^{5} \sim 10^{6}$ years
\citep{Salvador2017,Lebrun2013,Hamano2015} named \emph{magma ocean}
(MO) stage.

The aim of this study is twofold: first, knowing the evolution of
these atmospheres would better constrain the early habitability of a
given telluric planet (defined here as the presence or not of liquid
water on its surface) and second, the determination of the thermal
emission spectrum and the spectral reflectance of atmospheres of MO
planets would yield precious observational constraints upon MO
exoplanets. In this context, several studies have already been made,
generally using models coupled with various submodules (interior,
atmosphere, escape) interacting with each other through matter and/or
energy fluxes \citep{Hamano2015,Lupu2014,Lebrun2013}. However, some of
these models prescribe the albedo instead of computing it in a
self-consistent way. Considering that albedo plays a major role in the
thermal balance, and thus in the habitability and observable spectrum
of such planets, the present study aims to improve its computation.

We will first describe the models used and the improvements made in
Section~\ref{sec:model}, then we will present the results in
Section~\ref{sec:results} and finally we will discuss the limitations
of our model and suggests some solutions in Section~\ref{sec:discus}.

\section{Model description}
\label{sec:model}

The atmospheric model used here, developed initially in \citet{Marcq2012} and
recently updated in \citet{Marcqetal2017}, is a one-dimensional model
extending from the surface up to an altitude corresponding to $0.1$
Pa, taking into account radiative and convective processes in the
atmosphere. The atmosphere, assumed here to be composed of H$_2$O and CO$_2$, is divided into plane-parallel computational layers
discretized according to pressure coordinates.

\begin{figure*}
\centering
\includegraphics[scale=0.50]{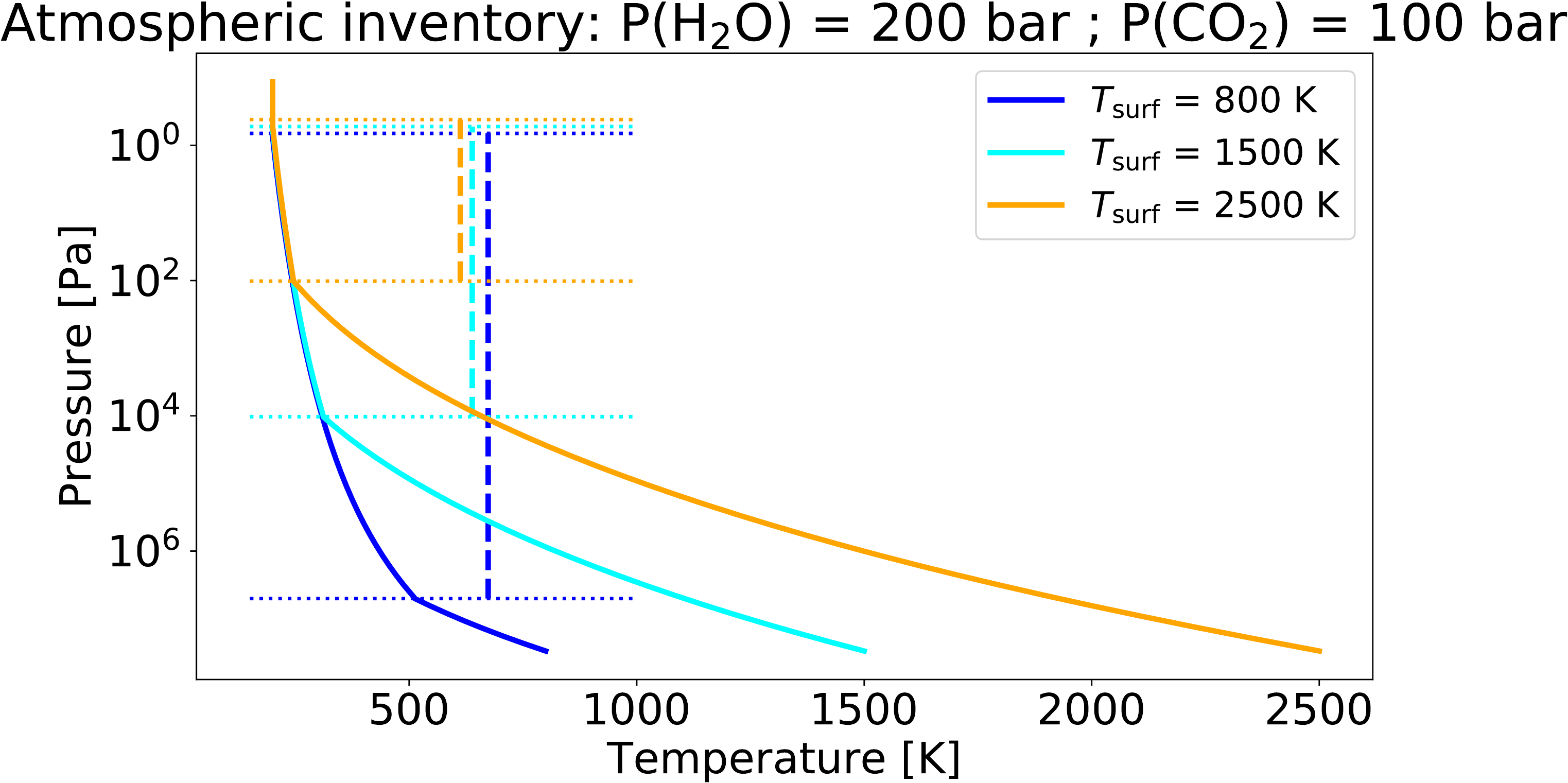}\\
\vfill
\includegraphics[scale=0.50]{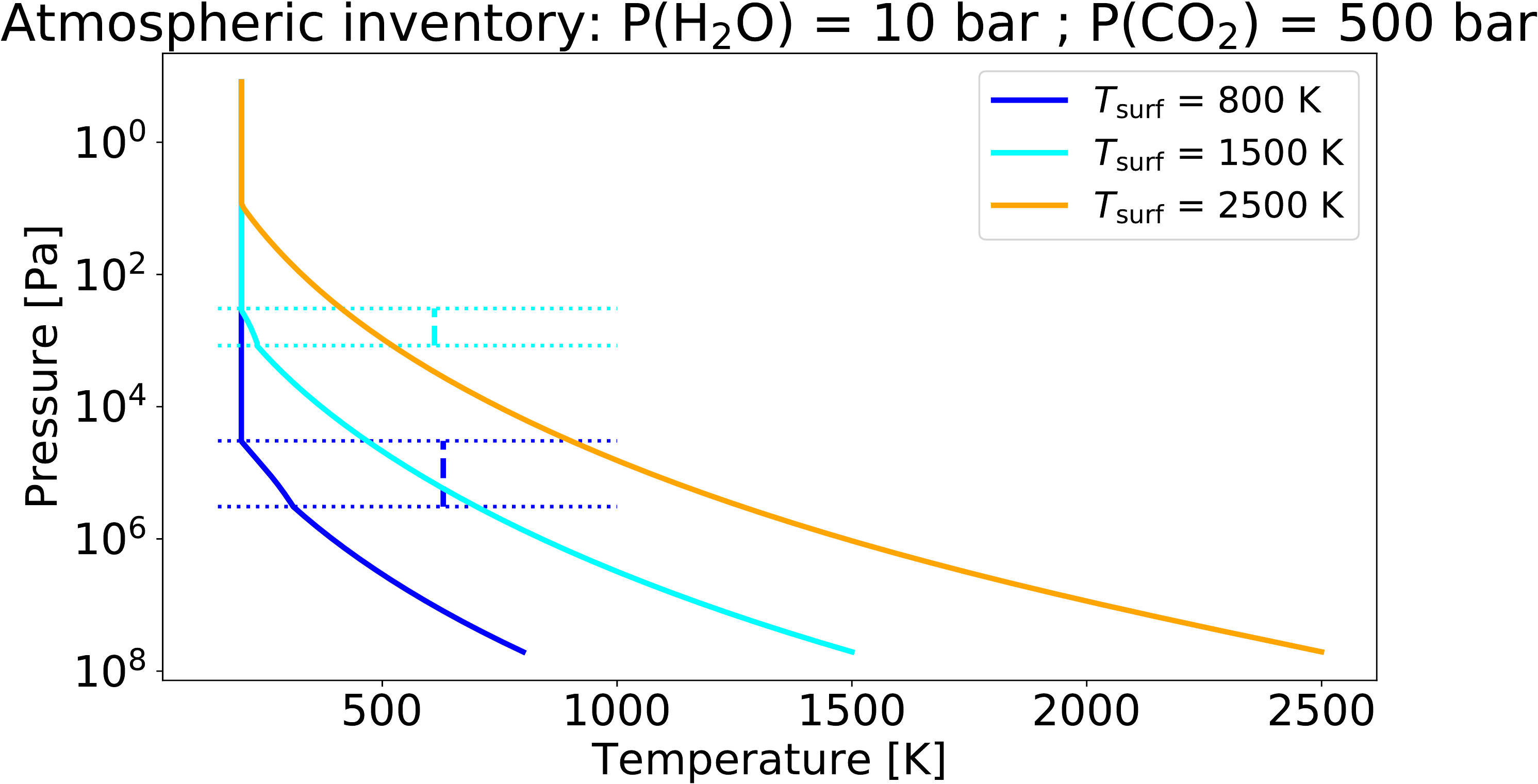}
\caption{Temperature profiles for a water-dominated atmosphere (up) and a CO$_2$-dominated atmosphere (down) at three surface temperatures using the model from \citet{Marcqetal2017}. The horizontal dotted lines show the moist/dry troposphere limits and the moist troposphere/mesosphere limits for the three surface temperatures assumed. Finally, the vertical dotted lines delimit the cloud layer vertical extent for each surface temperature.}
\label{fig:Profil-de-temperature}%
\end{figure*}

\begin{figure*}
\centering
\includegraphics[scale=0.50]{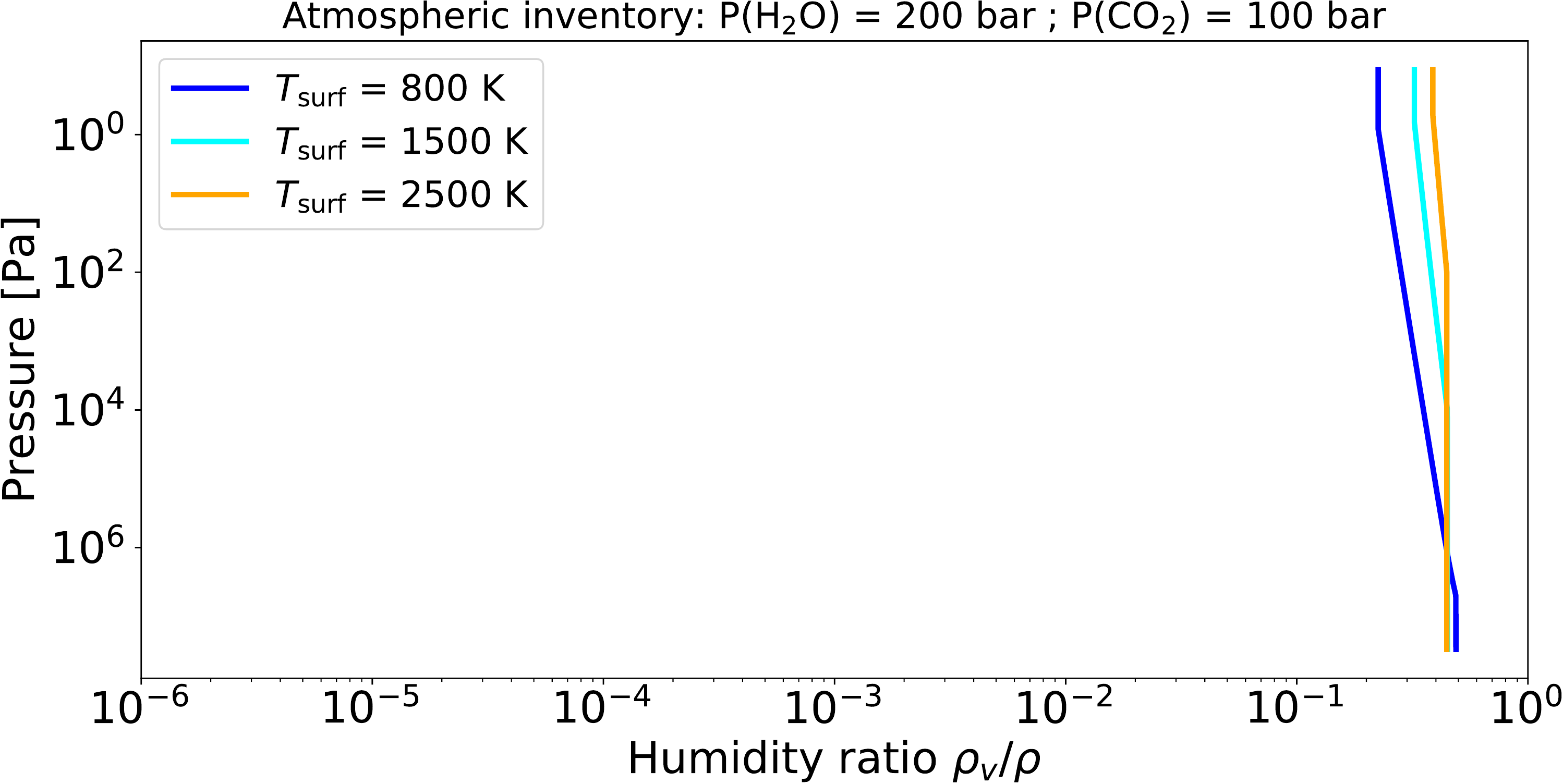}
\includegraphics[scale=0.50]{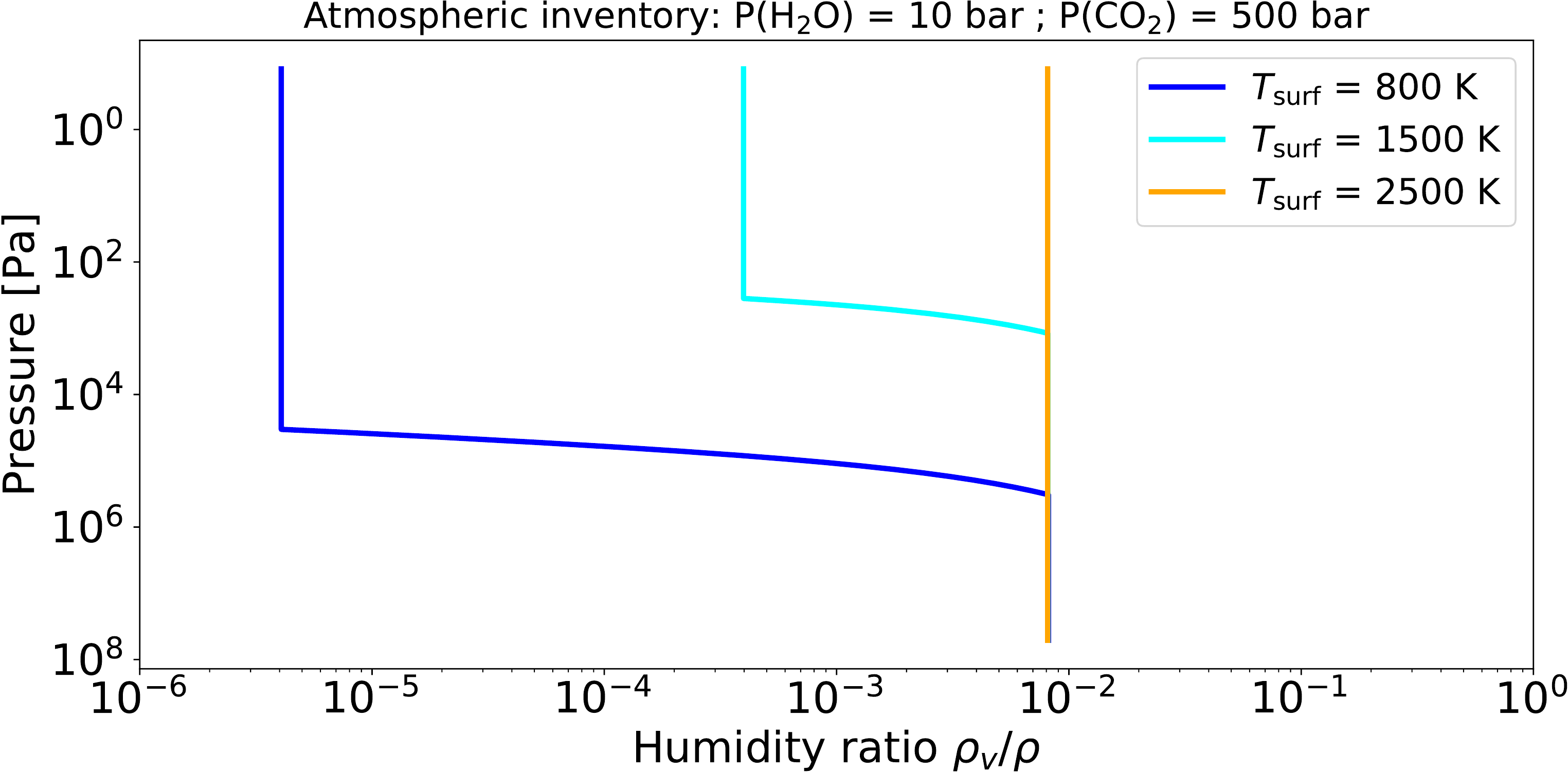}
\caption{Humidity profile for a water-dominated atmosphere (up) and a CO$_2$-dominated atmosphere (down) at three surface temperatures using the model from \citet{Marcqetal2017}.}
\label{fig:humidity_profile}%
\end{figure*}

The atmospheric profile used in this model is separated in three (at
most) physical layers \citep{Marcqetal2017}, displayed in Fig.~\ref{fig:Profil-de-temperature} and Fig.~\ref{fig:humidity_profile}. From the bottom up: (1) an
unsaturated troposphere, where heat transport is dominated by
convective processes. The humidity ratio H$_2$O/CO$_2$ is constant in this layer. Then (2) a moist troposphere, saturated in water
vapor, which is also a convective layer. It is in this layer that
clouds are prescribed, considered as terrestrial clouds from a
microphysical point of view (see \S\ref{sssec:cloud}). The dashed
vertical lines in Fig.~\ref{fig:Profil-de-temperature} represent the
vertical extent of this cloud layer for three different surface
temperatures. The humidity ratio decrases with increasing height in the moist troposphere as shown in Fig.~\ref{fig:humidity_profile}. Please take note that if the moist troposphere reaches the
ground, which happens whenever the partial surface pressure of H$_2$O
at the surface exceeds the saturation pressure and if we are below the critical point of water, then an ocean of
liquid water is expected to be formed on the surface of the
planet~\citep{Marcqetal2017} with the excess water content. Finally
(3), a purely radiative, isothermal mesosphere with a temperature
fixed to $T_0=200\,\mathrm{K}$ following other
studies~\citep{Lupu2014,Leconte2013} which have shown that even for
very hot surface temperatures, these atmospheres exhibit rather cool
temperatures at their top. In this layer, humidity remains constant (at a lower value: we have a cold trap) in the mesosphere. The analytical expressions for H$_2$O/CO$_2$ profiles are the same as in \citep{kasting1988}. We focused our study on magma ocean planets located far enough to be able to cool, that is far enough of their host stars so that they absorb less than $\sim 300\,\mathrm{W/m^2}$ (Nakajima's limit). In such a regime, most comparable studies find cool mesospheres~\citep{Lupu2014,Leconte2013}.

The radiative transfer is performed using the radiative standard solver \texttt{DISORT} \citep{StamnesEtAl1988} in a four-stream approximation in order to compute the thermal emission in 36 spectral intervals (given in Tab.
\ref{tab:spectral_intervals}). For this, we use the correlated-$k$ approach \citep{FuLiou1992}. We use 16 Gauss points in the $g$-space integration, where $g$ is the cumulated distribution function of the spectrally-resolved line and continuum absorption data in each of the 36 bands. In order to calculate the opacity of the planetary atmosphere in the thermal range, the model makes use of the line opacities and continuum opacities of mixtures of CO$_2$ and H$_2$O, where both gases can become dominant.

\begin{table*}
\hfill{}%
\begin{tabular}{|c|c|}
\hline 
Spectral band [$\upmu$m] & Spectral band [$\upmu$m] \tabularnewline
\hline 
0.29-0.30 & 1.13-1.20\tabularnewline
\hline 
0.30-0.35 & 1.20-1.31\tabularnewline
\hline 
0.35-0.40 & 1.31-1.43\tabularnewline
\hline 
0.40-0.45 & 1.43-1.56\tabularnewline
\hline 
0.45-0.50 & 1.56-1.69\tabularnewline
\hline 
0.50-0.55 & 1.69-1.86\tabularnewline
\hline 
0.55-0.60 & 1.86-2.02\tabularnewline
\hline 
0.60-0.65 & 2.02-2.20\tabularnewline
\hline
0.65-0.67 & 2.20-2.48\tabularnewline
\hline
0.67-0.69 & 2.48-2.66\tabularnewline
\hline 
0.69-0.75 & 2.66-2.92\tabularnewline
\hline 
0.75-0.78 & 2.92-3.24\tabularnewline
\hline 
0.78-0.84 & 3.24-3.58\tabularnewline
\hline 
0.84-0.89 & 3.58-4.01\tabularnewline
\hline 
0.89-0.96 & 4.01-4.17\tabularnewline
\hline 
0.96-1.04 & 4.17-4.55\tabularnewline
\hline
1.04-1.07 & 4.55-4.88\tabularnewline
\hline
1.07-1.13 & 4.88-5.13\tabularnewline
\hline
\end{tabular}\hfill{}

\caption{List of the 36 thermal infrared spectral bands used in the model.\label{tab:spectral_intervals}}
\end{table*}

\subsection{Model update }
\label{ssec:update}

The aim of this study is to determinate the Bond (or bolometric)
albedo of MO planets. For this, we extended the spectral range
of our model to optical and near UV domains. This implies some model
improvements and modifications which are detailed in the following subsections.

\subsubsection{Rayleigh scattering }

Due to new shorter wavelength range under study, Rayleigh scattering
opacity was added to the most recent version of the atmospheric
submodule~\citep{Marcqetal2017}. Rayleigh cross-sections of H$_2$O and CO$_2$ were assumed to follow a simple $\lambda^{-4}$ spectral dependency:
$\sigma_{\mathrm{Rayleigh}}(\lambda)=\sigma_{0}(\frac{\lambda_{0}}{\lambda})^{4}$
with $\sigma_{0}$ and $\lambda_{0}$ (given in Tab.
\ref{tab:cross_section}) so that spectral averaging within each of the
36 bands in both thermal (1$-$200 $\mu$m) and stellar
(0.2857$-$5.1282 $\mu$m) components could be computed
analytically. Although the spectral dependency of
$\sigma_{\mathsf{rayleigh}}(\lambda)$ with the wavelength is actually
more complex \citep{SNEEP2005293}, we used this simple analytical equation across the whole
spectrum. This choice is justified because cross-sections have been
estimated around a wavelength of $0.6$\,\textmu m, which is where the
details of Rayleigh scattering opacity are most relevant (see also
\S\ref{ssec:reflect}).

\begin{table*}
\hfill{}%
\begin{tabular}{|c|c|c|c|}
\hline 
Molecules & $\sigma_{0}$ {[}cm\texttwosuperior /molecule{]} & $\lambda_{0}$ {[}\textmu m{]} & Bibliographical references\tabularnewline
\hline 
\hline 
H$_2$O & $2.5\,10^{-27}$ & $0.6$ & \citet{KopparapuRamirezKastingEtAl2013}\tabularnewline
\hline 
CO$_2$ & $1.24\,10^{-26}$ & $0.532$ & \citet{SNEEP2005293}\tabularnewline
\hline 
\end{tabular}\hfill{}

\caption{Table showing the Rayleigh scattering cross-sections and corresponding wavelengths for the two molecules used in our study.\label{tab:cross_section}}
\end{table*}

\subsubsection{Gaseous line absorption}
\label{sssec:line}

We used here new line opacity data supplementing those of~\citet{Marcqetal2017}.
Opacity caused by the absorption of H$_2$O and CO$_2$ in the atmosphere has been computed using $kspectrum$~\citep{Eymet2016} to yield high-resolution line-by-line spectra. We used the HITRAN2012 database for the H$_2$O and CO$_2$ line intensities and parameters \citep{Rothman2013}. Additionally, we incorporated here the half-width at half maximum of H$_2$O lines broadened by CO$_2$ ($\gamma^{\text{H}_2\text{O}-\text{CO}_2}$) and CO$_2$ lines broadened by H$_2$O ($\gamma^{\text{CO}_2-\text{H}_2\text{O}}$), as well as the  corresponding temperature dependence exponents ( $n^{\text{H}_2\text{O}-\text{CO}_2}$ and $n^{\text{CO}_2-\text{H}_2\text{O}}$ ), based on references~\citep{Brown2007,Sung2009,Gamache2016,Delahaye2016}. More details can be found in \citet{Turbet2017LPI} and \citet{Tran2018}.
High resolution spectra were then computed using $kspectrum$ \citep{Eymet2016}, for a given temperature/pressure/composition grid, detailed in
Tab.~\ref{tab:Tableaux_grilles_opacite}. 
\\ Given the range of temperatures/pressures probed here, and given the importance of continuum absorption in the dense atmosphere discussed here, we believe that -- as of 2018 -- the HITRAN database is more reliable for the calculations of absorption coefficients used in the present work than any other existing database, including ExoMol.
Indeed, ExoMol cross-sections were calculated assuming H$_2$ and He-dominated atmospheres, i.e. assuming line broadening by H$_2$/He, whereas, in the present study, we used self-broadening coefficients $\gamma^{\text{CO}_2-\text{CO}_2}$/$\gamma^{\text{H}_2\text{O}-\text{H}_2\text{O}}$ provided by HITRAN and foreign broadening coefficients $\gamma^{\text{CO}_2-\text{H}_2\text{O}}$/$\gamma^{\text{H}_2\text{O}-\text{CO}_2}$ (and associated temperature dependency exponents $n^{\text{H}_2\text{O}-\text{CO}_2}$/$n^{\text{CO}_2-\text{H}_2\text{O}}$) described above. Plus, some band lines are missing from the ExoMol database. For instance, CO$_2$ band lines around 1300-1400~cm$^{-1}$ are missing from ExoMol. More generally, we recall that the widely used HITRAN database is based on both experiments and calculations, whereas ExoMol cross-sections/line lists are based on theoretical calculations only.

\begin{table*}
\hfill{}%
\begin{tabular}{|c|c|c|}
\hline 
\multicolumn{3}{|c|}{Temperature/pressure/composition grid used in this study}\tabularnewline
\hline 
Pressure (Pa) & Temperature (K) & Volume mixing ratio (H$_2$O/(CO$_2$+H$_2$O)) \tabularnewline
\hline 
\hline 
$10^{7}$ & $900$ & $10^{-7}$\tabularnewline
\hline 
$10^{6}$ & $700$ & $10^{-6}$\tabularnewline
\hline 
$10^{5}$ & $500$ & $10^{-5}$\tabularnewline
\hline 
$10^{4}$ & $420$ & $10^{-4}$\tabularnewline
\hline 
$10^{3}$ & $360$ & $10^{-3}$\tabularnewline
\hline 
$10^{2}$ & $300$ & $10^{-2}$\tabularnewline
\hline 
$10^{1}$ & $250$ & $10^{-1}$\tabularnewline
\hline 
$10^{0}$ & $200$ & $3\times10^{-1}$\tabularnewline
\hline 
$10^{-1}$ &  & $7\times10^{-1}$\tabularnewline
\hline 
 &  & $9\times10^{-1}$\tabularnewline
\hline 
 &  & $9.9\times10^{-1}$\tabularnewline
\hline 
\end{tabular}\hfill{}

\caption{Table showing the temperature/pressure/composition grid used to calculate gaseous absorption. \label{tab:Tableaux_grilles_opacite}}
\end{table*}

\subsubsection{Continuum absorption}
\label{sssec:continua}

As we study dense atmospheres, we have to take into account collision
induced absorption, dimer absorptions and far line wing absorptions. These absorptions are especially relevant to assess the transparency of possible spectral windows wherever line opacity is weak. 

Far line wings of CO$_2$ (CO$_2$-CO$_2$ and CO$_2$-H$_2$O) were computed using the $\chi$-factor approach as in \citet{Tran2018}. This is an empirical correction of the Lorentzian line shape adjusted to laboratory measurements. CO$_2$-CO$_2$ collision-induced and dimer absorptions were computed based on references \citep{Gruszka1997,Baranov2004,Stefani2013}. 

H$_2$O-H$_2$O continuum was taken into account using the MT$\_$CKD~3.0 database \citep{Mlawer2012}, from 0 to 20,000~cm$^{-1}$. MT$\_$CKD databases are available on \url{http://rtweb.aer.com/}. H$_2$O-CO$_2$ continuum was calculated with the line shape correction functions of \citet{Ma1992} using line positions and intensities from the HITRAN2012 database \citep{Rothman2013}, with a cut-off distance at 25~cm$^{-1}$, and from 0 to 20,000~cm$^{-1}$. The temperature dependence of the continuum was empirically derived using data digitized from \citet{POLLACK19931}. More details can be found in \citet{Turbet2017LPI} and \citet{Tran2018}.

\subsubsection{Cloud optical properties}
\label{sssec:cloud}

In this updated model, we take into account the cloud opacity as in \citet{Marcqetal2017}. We take into account the wavelength dependency of cloud opacity (see Tab.~\ref{tab:parametresnuages}) according to approximated Mie theory
calculations from \citet{kasting1988} and assuming spherical H$_2$O
droplets with a mean~5~\textmu m radius. 

Following~\citet{kasting1988}, the mass loading of these Earth-like clouds is
$\rho_{\mathrm{clouds}}=4\,10^{-4}\rho_{\mathrm{gases}}$. Single scattering phase function is assumed to follow the analytical Henyey-Greenstein expression with the asymmetry parameter $g$ given by~\citet{kasting1988}.

\begin{table*}
\hfill{}%
\begin{tabular}{|c|c|c|c|c|}
\hline 
Wavelength $\lambda$ [\textmu m] & $\lambda < 2$ & $2 \leq \lambda < 10$ & $10 \leq \lambda < 20$ & $20 \leq \lambda$ \tabularnewline
\hline 
\hline 
Single-scattering albedo $\varpi_0$ & $1$ & $1.24 \lambda^{-0.32}$ & $1.24 \lambda^{-0.32}$ & $1.24\lambda^{-0.32}$\tabularnewline
\hline 
Extinction efficiency $Q_{\mathrm{ext}}$ & $1$ & $1$ & $1$ & $3.26\lambda^{-0.4}$\tabularnewline
\hline 
Asymmetry factor $g$ & $0.85$ & $0.85$ & $1.4\times\lambda^{-0.22}$ & $1.4\times\lambda^{-0.22}$\tabularnewline
\hline 
\end{tabular}\hfill{}

\caption{Optical properties of clouds.\label{tab:parametresnuages}}
\end{table*}

\section{Results}
\label{sec:results}

\subsection{Spectral reflectance \& albedo}
\label{ssec:reflect}

\begin{figure*}
\includegraphics[scale=0.55]{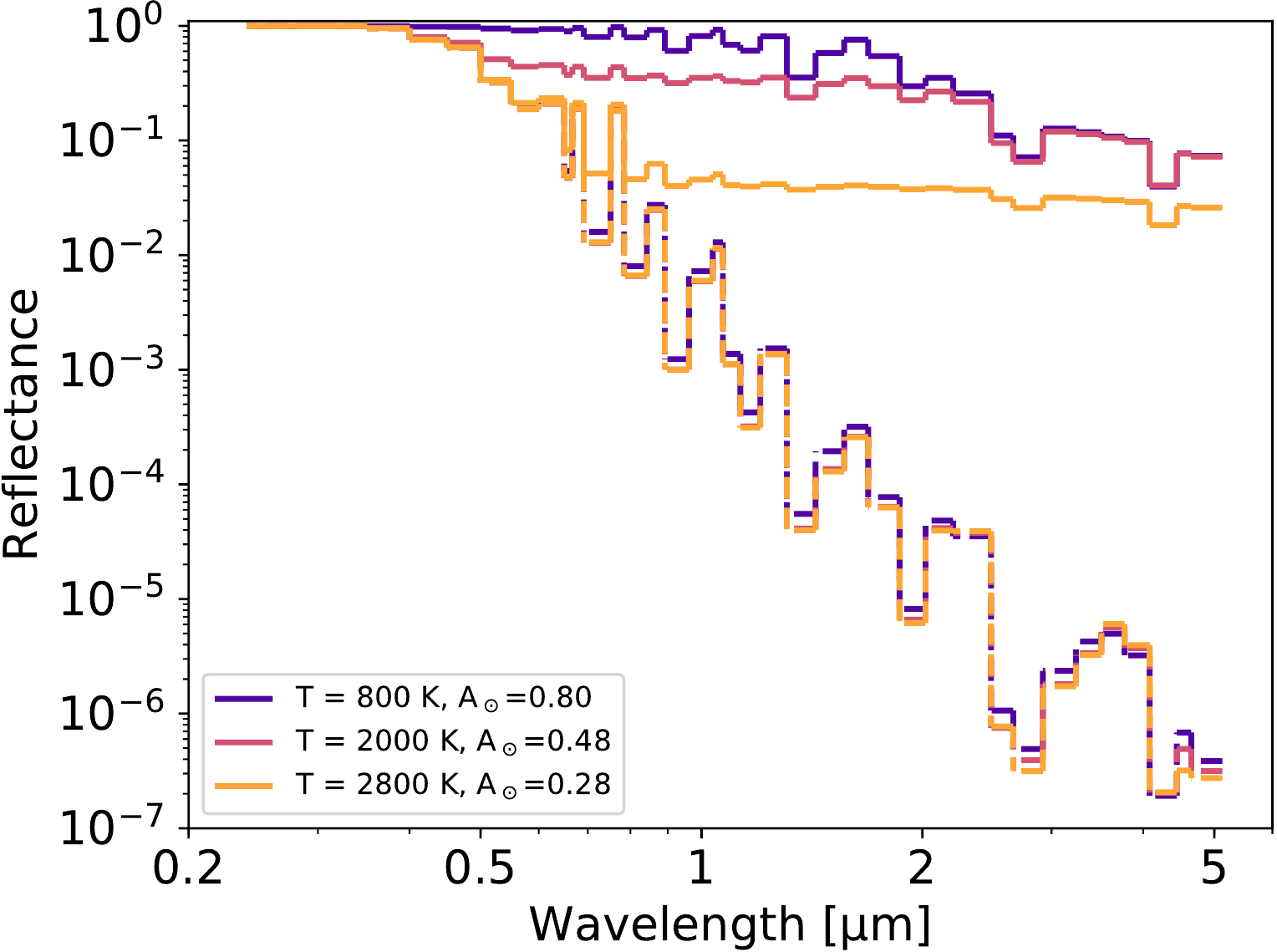}\hfill{}\includegraphics[scale=0.55]{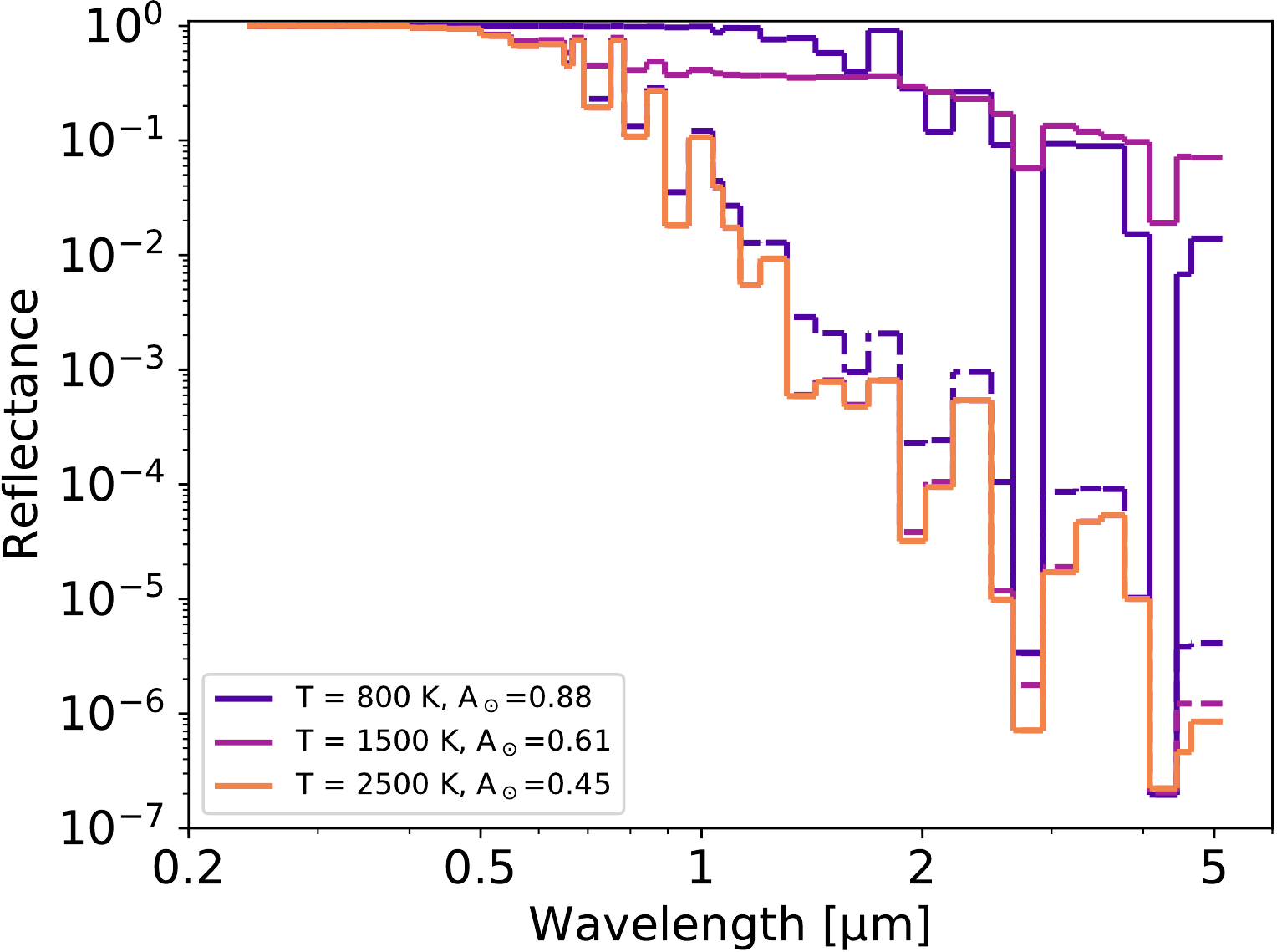}
\caption{Spectral reflectance with clouds (solid curve) and without clouds (dashed
curve), for three different surface temperatures. For spectral reflectance with clouds, the corresponding Bond albedo for a Sun-like star is indicated. The sharp variations in reflectance are due to molecular absorption bands, such at the 2.7 \textmu m (H$_2$O) and the 4.3 \textmu m (CO$_2$) features. Atmospheric inventory : (left) $P_{\mathrm{H_2O}} = 200\,\mathrm{bar}$,
  $P_{\mathrm{CO_2}}=100\,\mathrm{bar}$. (right) $P_{\mathrm{H_2O}} = 10\,\mathrm{bar}$, $P_{\mathrm{CO_2}}=500\,\mathrm{bar}$. \label{fig:Courbe-de-reflectance}}
\end{figure*}

Now that we have taken Rayleigh scattering and atmospheric gaseous
opacities into account as well as the Mie scattering from clouds, we can calculate the
spectral reflectance of these atmospheres, as shown in Fig.~\ref{fig:Courbe-de-reflectance}. For this, we use the solver \texttt{DISORT} (\citealp{StamnesEtAl1988}) and assume a mean isotropic illumination at the top of the atmosphere. 
We can see on Fig.~\ref{fig:Courbe-de-reflectance} that the spectral reflectance with or without clouds is very high under $0.5$~\textmu m due to very efficient Rayleigh scattering at shorter wavelengths. 

Beyond $0.5\,$\textmu m, the behavior of the spectral reflectance strongly depends on the surface temperature of the planet and whether clouds are taken into account or not. Without clouds in the atmosphere, there is a competition between Rayleigh scattering and gaseous absorption. Overall, Rayleigh scattering fails to compensate for H$_2$O and CO$_2$ absorption in the near infrared resulting in a rapid collapse of spectral reflectance at smaller wavenumbers, as shown by the dashed lines in Fig.~\ref{fig:Courbe-de-reflectance}. Nevertheless, there are still local reflectance maxima where gaseous absorption is very low (spectral windows).

When clouds are taken into account, we no longer observe a collapse of the spectral reflectance towards smaller wavenumbers, but a more modest decrease by one order of magnitude. This is due to the contribution of Mie scattering to spectral reflectance. However, we then observe two distinct behaviors of the spectral reflectance at smaller wavenumbers. Firstly, it depends primarily on the surface temperature: the higher the surface temperature for a given atmosphere, the thinner the cloud layer and its reflectance is reduced. Secondly, for any given surface temperature, the spectral reflectance is decreasing with decreasing short wavenumber. Indeed, Mie scattering becomes less and less efficient at smaller wavenumber (see Tab.~\ref{tab:parametresnuages}). Also, we can note some spectral structure in the spectral reflectance, which is due to absorption bands of H$_2$O and CO$_2$ in the near infrared even in the presence of clouds -- typically water absorption bands at about $2.7\,$\textmu m and carbon dioxide absorption at about $4.3\,$\textmu m (see Fig.~\ref{fig:Courbe-de-reflectance}).

Once the spectral reflectance is computed, the resulting Bond albedo
can be calculated by weighting the reflectivity of each band of the 36~bands according
to the stellar emission, assuming that it behaves like a blackbody of
effective temperature $T_*$ at the studied wavelengths. This yields
the following formula: \[
A_{B}(R;T_*)=\frac{\int_{0}^{\infty}R(\tilde{\nu}) B_{\tilde{\nu}}(T_{*})\,d\tilde{\nu}}{\int_{0}^{\infty}B_{\tilde{\nu}}(T_{*})\,d\tilde{\nu}}=\frac{\pi \int_{0}^{\infty}R(\tilde{\nu})B_{\tilde{\nu}}(T_{*})d\tilde{\nu}}{\sigma T_{*}^{4}}
\] where $R(\tilde{\nu})$ stands for the spectral reflectance as a function
of wavenumber, $B_{\tilde{\nu}}(T_{*})$ the Planck as a function of
wavenumber and stellar temperature and $\sigma$ the
Stefan-Boltzmann constant.

\subsection{Albedo behavior with respect to surface temperature}

\begin{figure*}
\hfill{}\includegraphics[scale=0.80]{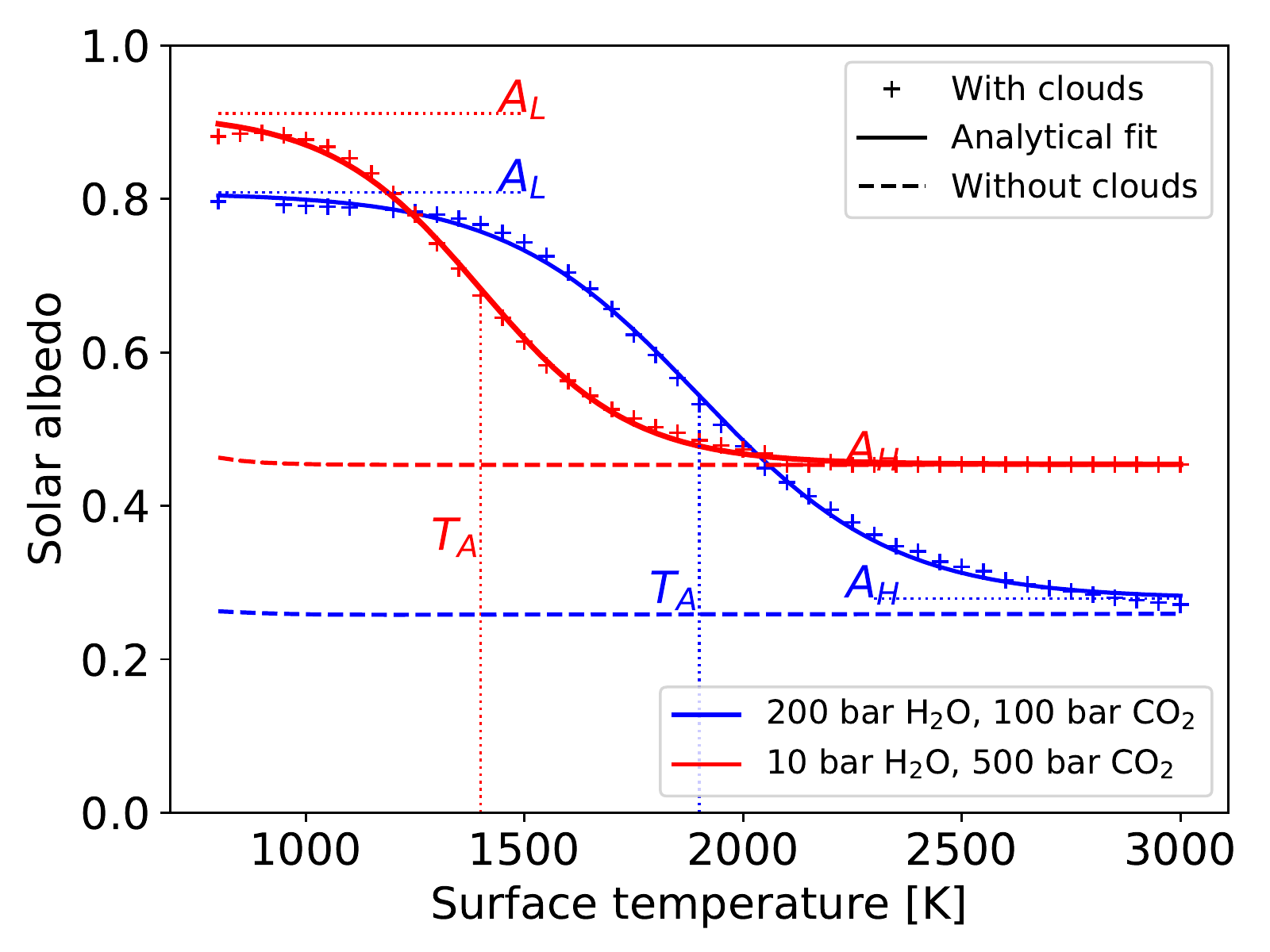}\hfill{}
\caption{Bond albedo as a function of surface temperature as computed
  by our model (crosses) and our analytic fit (solid) for two different atmospheres, H$_2$O-dominated (blue) and CO$_2$-dominated (red)  illuminated by the Sun
  ($\mathrm{T}_*=5750\,\mathrm{K}$). The above plot indicates that, the planet has a high albedo at low surface temperatures due to Mie scattering dominating the gas absorption. At high surface temperatures, the albedo reaches asymptotic value of 0.25, consistent with \citet{Hamano2015}, due to dominance of the gas absorption. Intermediate temperatures result in decreasing albedo due to the shrinkage of the cloud deck, as shown in Fig. 1. \label{fig:Fit_albedo}}
\end{figure*}

\begin{figure*}
\hfill{}\includegraphics[scale=0.80]{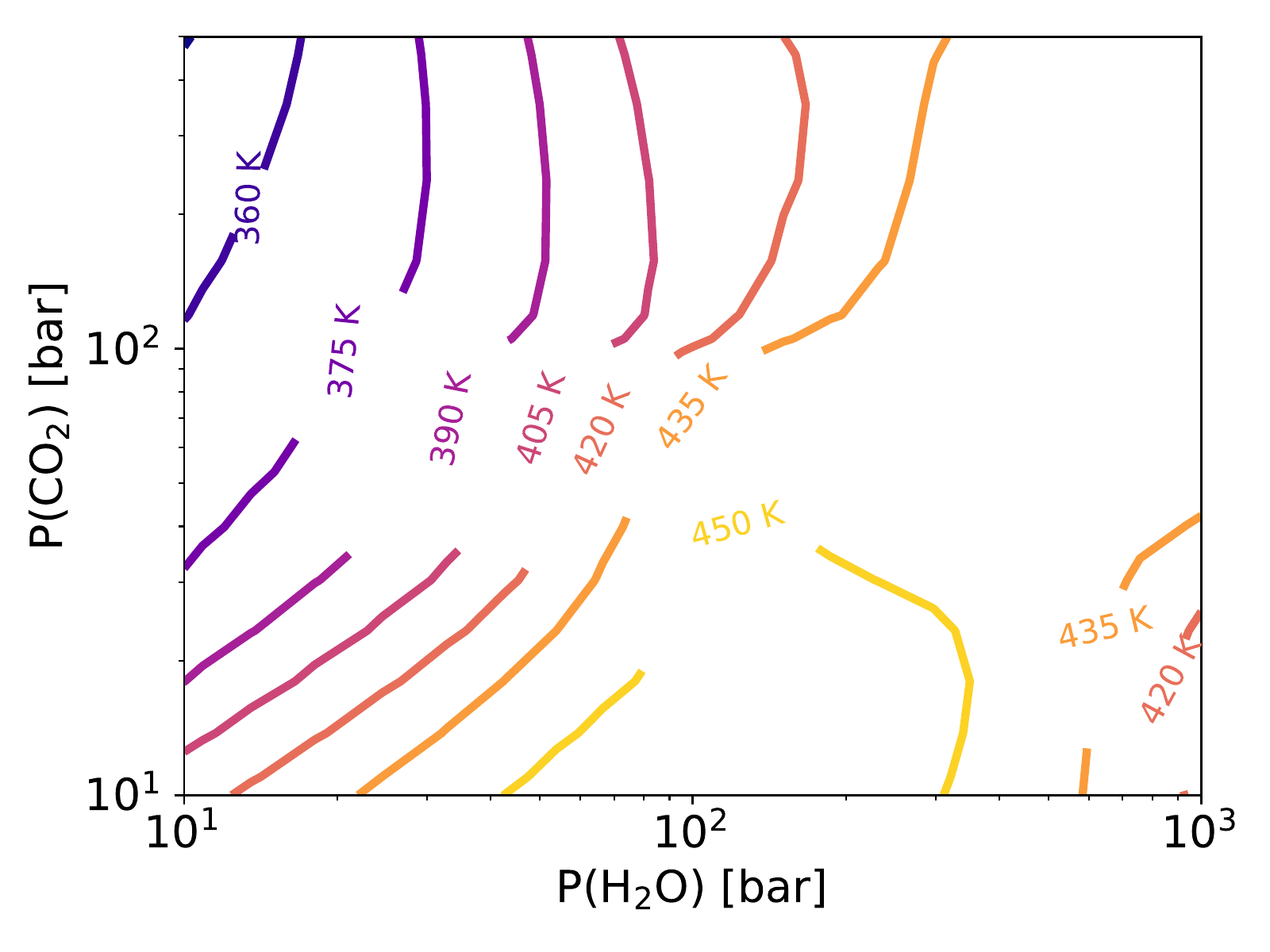}\hfill{}
\caption{Contour plot of the temperature interval width $T_{\mathrm{scale}}$ where the albedo transition occurs with
  respect to the partial pressures $\mathrm{P}_{\mathrm{surf}}$(H$_2$O) and
  $\mathrm{P}_{\mathrm{surf}}$(CO$_2$) for a planet around by the Sun
  ($\mathrm{T}_*=5750\,\mathrm{K}$). \label{fig:Tscale}}
\end{figure*}

The typical behavior of the Bond albedo with respect to the surface
temperature is shown in Fig.~\ref{fig:Fit_albedo} for a MO planet with an atmospheric composition dominated by water vapor, and taking clouds into account. Our model is unidimensional, therefore the clouds cover the entire planet. 
This seems reasonable to assume that such MO planets are either totally covered by clouds, as Venus today, or either totally cloudless, because these atmospheres are dominated by their internal heat flux. It has actually been demonstrated with a 3-Dimensions Global Climate Model that a complete cloud cover would be produced following very large meteoritic impact events \citep{Turbet2017impact}, while the thermal budget of the atmosphere is dominated by the infrared cooling of the atmosphere of the planet.
More generally, the internal heat flux of MO planets should not strongly depend on local time or latitude, so that we expect much more horizontally uniform atmospheres than for atmospheres dominated by the stellar flux. This Fig.~\ref{fig:Fit_albedo} reveals three main features:

\begin{enumerate}
\item For relatively low surface temperatures, the planet exhibits a
  very high and temperature-independent albedo (about $0.8-0.9$
  here). This is due to the Mie scattering in the clouds completely
  dominating the gaseous absorption. Backscattered photons travel a short optical path in the atmosphere and are then less absorbed. This situation is comparable to Venus, which also exhibits a very high albedo due to its thick uniform cloud cover.
\item For intermediate surface temperatures, the albedo decreases
  gradually with increasing surface temperature. Indeed, when the
  surface temperature increases, the cloud layer vertical extent and more importantly, the cloud layer optical depth (in the moist troposphere) shrinks (see Fig.~\ref{fig:Profil-de-temperature} and Fig.~\ref{fig:humidity_profile}). The cloud base moves upwards, at pressure levels where the mass loading of these clouds is much lower. The Mie scattering becomes less efficient compared to the gaseous absorption, mostly unaffected by changes in cloud content.
\item For very high surface temperatures, the albedo decreases towards
  an asymptotic value (at $0.25$ here), which corresponds to the
  albedo of a cloudless atmosphere. In this domain, relatively
  inefficient Rayleigh scattering is the only counterpart to gaseous
  absorption since Mie scattering by the clouds has become
  negligible. The value of $0.25$ is rather in good agreement with the work
  of~\citet{Hamano2015} who found an albedo of $0.22$ for such high
  temperatures and similar atmospheric composition.
\end{enumerate}
The qualitative trends described above are independent of the atmospheric composition for all the 42 compositions that we modeled (see Tab.~\ref{tab:Grille-en-pression}). An illustration of this is shown in Fig.~\ref{fig:Fit_albedo} for two different atmospheres, H$_2$O-dominated (blue) and CO$_2$-dominated (red).
\\In order to investigate the effects of varying e.g. the atmospheric
composition, we sought to extract some meaningful
parameters from the curves $\mathsf{A_{B}}=f(\mathsf{T_{surf})}$. The
empirical analytical fit:
\[ A_{B}\left(T_{\mathrm{surf}}\right) = \left( \frac{A_L-A_H}{2} \right) \tanh \left( \frac{T_A-T_{\mathrm{surf}}}{T_{\mathrm{scale}}} \right) + \frac{A_H+A_L}{2} \]

yields a surprisingly good fit (Fig.~\ref{fig:Fit_albedo}) of our
modeling. Here, $\mathrm{T}_\mathrm{A}$ defines our \emph{albedo transition temperature}
between the two regimes described above. In other words, $T_A$ is the surface temperature yielding an intermediate albedo $\left( \frac{A_L-A_H}{2} \right)$, where $A_H$ and $A_L$ respectively
stand for the high- and low-temperature asymptotic values of the
albedo. Finally $T_{\mathrm{scale}}$ corresponds to the
temperature range where the albedo transition
occurs. $T_{\mathrm{scale}}$ is the extent of the surface temperature domain centered on $T_A$ where the albedo lies in the 12-88 \% interval between $A_L$ and $A_H$. $T_{\mathrm{scale}}$ appears to be on the order of $400~\pm~40\,\mathrm{K}$ for all investigated atmospheres, as shown in Fig.~\ref{fig:Tscale} .


\subsection{Albedo vs. atmospheric inventory}
\label{ssec:composition}

As stated in the introduction, the atmospheres considered in this
model of MO planets are only composed of H$_2$O and CO$_2$
\citep{Lupu2014}. It is possible that these atmospheres were dominated
by other species, and our results indeed do show that atmospheric
inventory has an impact on their albedo. To analyse this impact,
we ran our model for 42 compositions according to the composition grid in Tab.~\ref{tab:Grille-en-pression}.

\begin{table*}
\hfill{}%
\begin{tabular}{|c|c|c|c|c|c|c|c|}
\hline $P_{\mathrm{surf}}$(H$_2$O) [bar] & $10$ & $20$ & $50$ & $100$ &
$200$ & $500$ & $1000$\tabularnewline
\hline 
$P_{\mathrm{surf}}$(CO$_2$) [bar] & $10$ & $20$ & $50$ & $100$ & $200$ & $500$ &
N/A\tabularnewline
\hline 
$T_{\mathrm{surf}}$ [K] & \multicolumn{7}{l|}{$800$--$3000$ ; step of $50$}
\tabularnewline
\hline
\end{tabular}\hfill{}
\caption{$T_{\mathrm{surf}}$, H$_2$O and CO$_2$ pressure grid used to simulate $7 \times 6 = 42$ different atmospheric inventories and surface temperature grid (45 points). \label{tab:Grille-en-pression}}
\end{table*}

We then investigated how our previously derived quantities ($\mathrm{T}_\mathrm{A}$,
$\mathrm{A}_\mathrm{H}$, $\mathrm{A}_\mathrm{L}$) changed with the atmospheric
inventory. Fig.~\ref{fig:Contour_TA} shows transition temperature
albedo's behavior $\mathrm{T}_\mathrm{A}$ according to the H$_2$O and the CO$_2$ partial
pressures. This figure shows that this transition temperature occurs
at higher values when atmospheric water content increases. Indeed, for
a given surface temperature, the more water available the optically
thicker the cloud layer (as shown in Fig.~\ref{fig:Profil-de-temperature} and Fig.~\ref{fig:humidity_profile}), so that the albedo remains high over a wider
range of surface temperatures, up to $2400\,\mathrm{K}$ for our
largest investigated water content value ($1000\,\mathrm{bar}$). On
the other hand, CO$_2$ has a marginal opposite effect, lowering the
humidity and therefore cloud opacity, all other parameters kept the
same. This behavior is reminiscent of the behavior of the threshold
temperature $\mathrm{T}_{\varepsilon}$ defined by \citet{Marcqetal2017} as the surface
temperature $T_\mathrm{s}$ where the effective emissivity $\varepsilon(\mathrm{T}_\mathrm{s}) = \mathrm{OLR}(\mathrm{T}_\mathrm{s})/\sigma {\mathrm{T}_{\mathrm{s}}}^4$ reaches its minimal value. This threshold temperature determines the transition between the low temperature regime $T_{\mathrm{s}} < T_{\varepsilon}$ where OLR is relatively low and constant ("runaway" regime) and the high temperature regime $T_{\mathrm{s}} > T_{\varepsilon}$ where OLR increases significantly with increasing $T_{\mathrm{s}}$ ("post-runaway" regime).

\begin{figure*}
\hfill{}\includegraphics[scale=0.37]{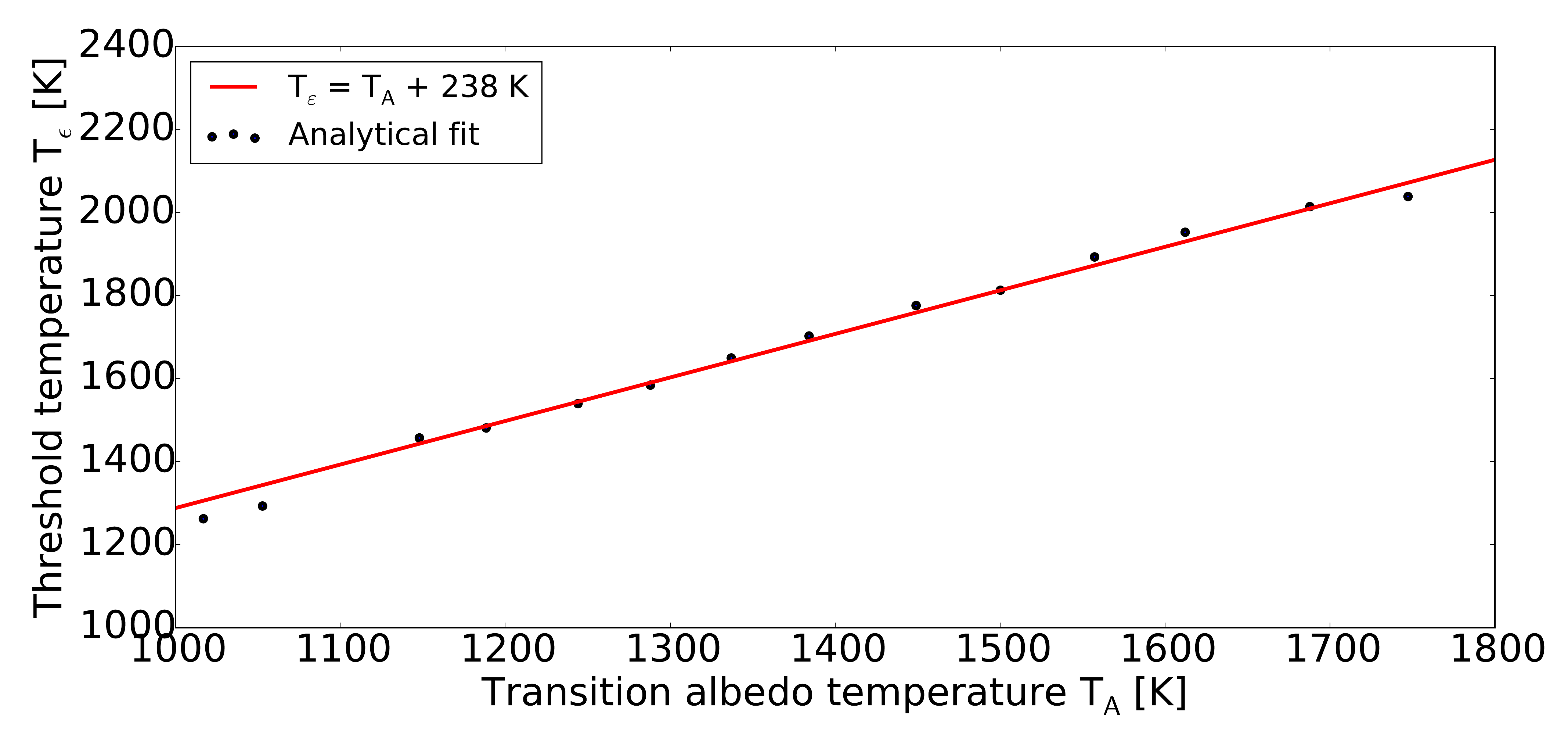}\hfill{}
\caption{Threshold temperature $T_{\varepsilon}$ with respect to the
  albedo transition temperature $T_A$ (black dots) and its linear fit
  (solid red).\label{fig:Temp_vs_temp}}
\end{figure*}

We therefore plotted $\mathrm{T}_{\varepsilon}$ vs. $\mathrm{T}_\mathrm{A}$ on Fig.~\ref{fig:Temp_vs_temp}. This figure shows that these two
temperature are actually the same within a constant offset
($\mathrm{T}_{\varepsilon}-\mathrm{T}_{A} \simeq 240\,\mathrm{K}$), which testifies to
the fact that the vertical structure of the atmosphere determines both
its thermal emission and its albedo. As the surface temperature
increases, clouds are becoming thinner, the albedo decreases and the
upper layers become hot enough for the thermal radiation to
substantially increase. Conversely, relatively cool MO
planets harbor both optically thick clouds (yielding a high albedo)
and a small outgoing thermal flux, close to $280\,\mathrm{W/m^2}$ when
water vapor dominates the NIR spectrum, see~\citet{Marcqetal2017}.

\begin{figure*}
\hfill{}\includegraphics[scale=0.8]{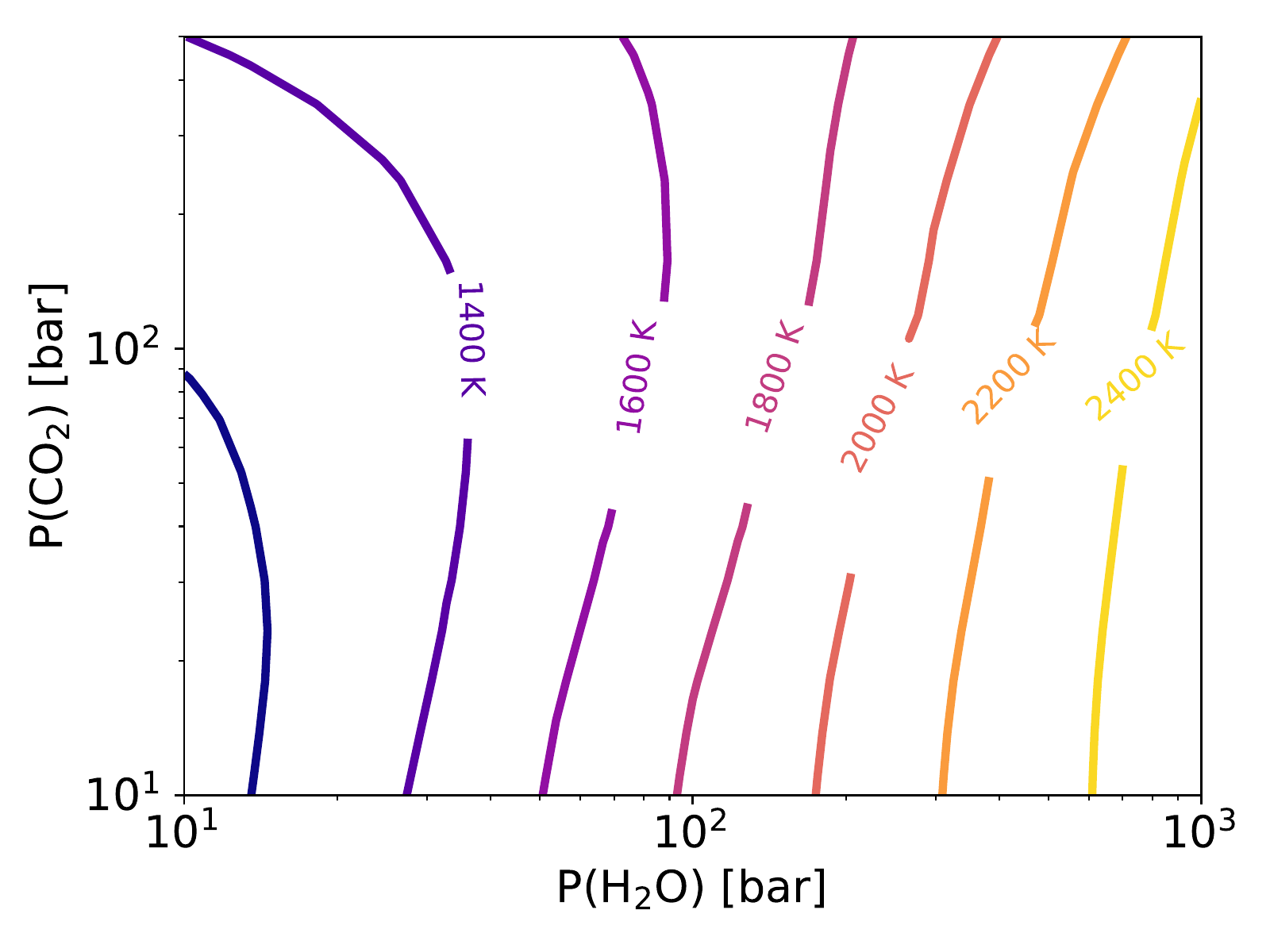}\hfill{}

\caption{Contour plot of the albedo transition temperature $T_A$ with
  respect to the partial pressures $P_{\mathrm{surf}}$(H$_2$O) and
  $P_{\mathrm{surf}}$(CO$_2$) for a planet around by the Sun
  ($T_*=5750\,\mathrm{K}$).\label{fig:Contour_TA}}
\end{figure*}

\begin{figure*}
\hfill{}\includegraphics[scale=0.8]{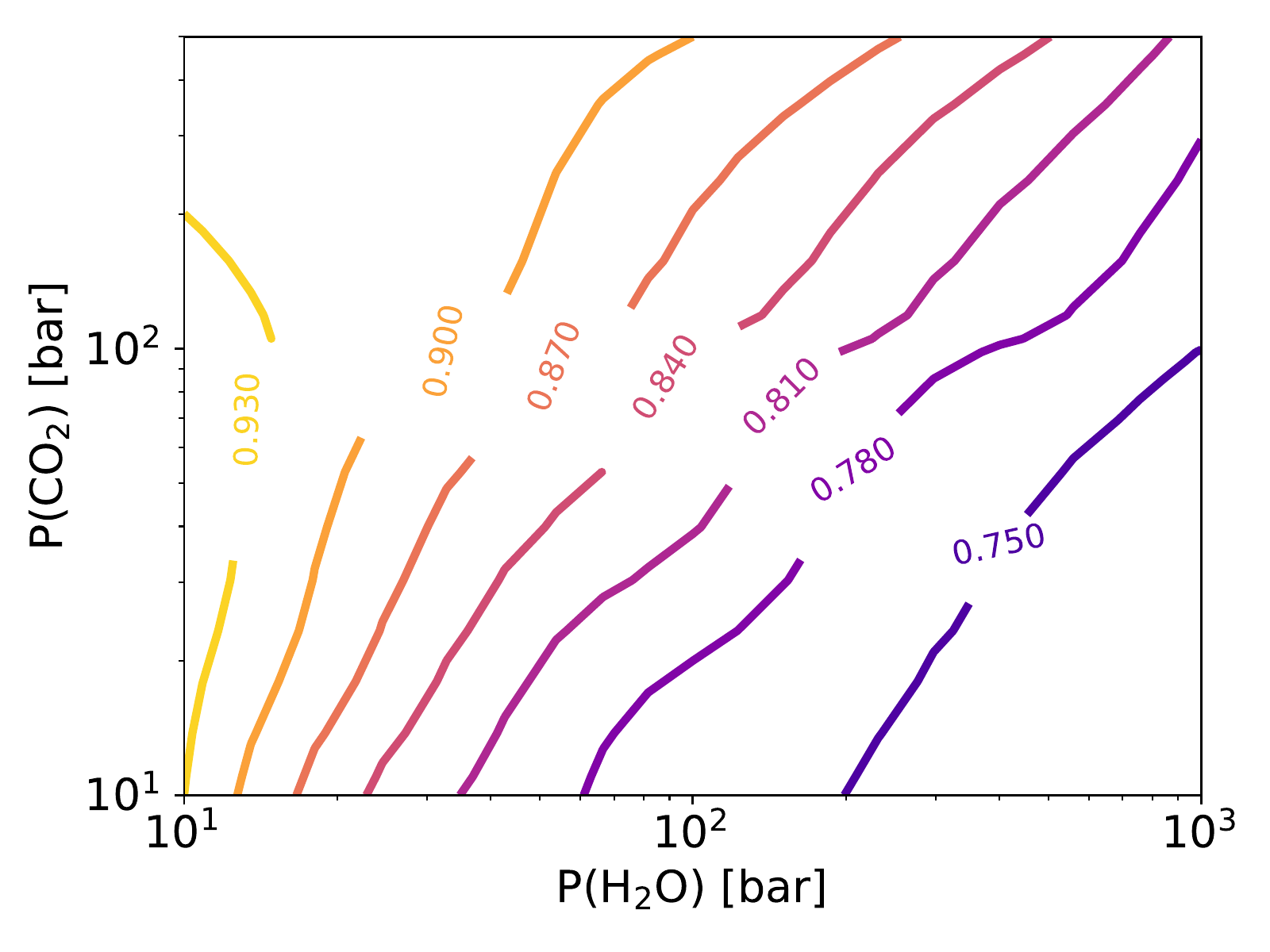}\hfill{}

\caption{$A_L$ contour plot with respect to H$_2$O partial pressure
  $P_{\mathrm{surf}}$(H$_2$O) and CO$_2$ partial pressure
  $P_{\mathsf{surf}}$(CO$_2$), taking into account the effects of
  cloud and assuming a Sun-like star
  ($T_*=5750\,\mathrm{K}$).\label{fig:Contour_800K}}
\end{figure*}

For surface temperatures smaller than the albedo transition
temperature $T_A$, the asymptotic low temperature albedo $A_L$
exhibits a fairly simple behavior. On one hand,
Fig.~\ref{fig:Contour_800K} shows that for a given CO$_2$ partial
pressure, the albedo increases when the pressure in H$_2$O
decreases. On the other hand, for a given H$_2$O partial pressure, the
albedo increases when the partial pressure in CO$_2$ increases. These
two statements reflect the same physical phenomenon, namely that
H$_2$O is a stronger absorber of stellar radiation than CO$_2$,
especially in the near IR. Drier atmospheres above the cloud top
absorb less, therefore increasing the albedo (since the single
scattering albedo of cloud particles is assumed to be close to unity
for $\lambda < 2\,$\textmu m). The very high values for $A_L$ for dry atmospheres are however not very robust, since any trace contaminant in the clouds would significantly lower the albedo (as it is indeed the case with Venus, whose Bond albedo of $0.7$ is lower than we would assume from Fig.~\ref{fig:Contour_800K}).

\begin{figure*}
\hfill{}\includegraphics[scale=0.8]{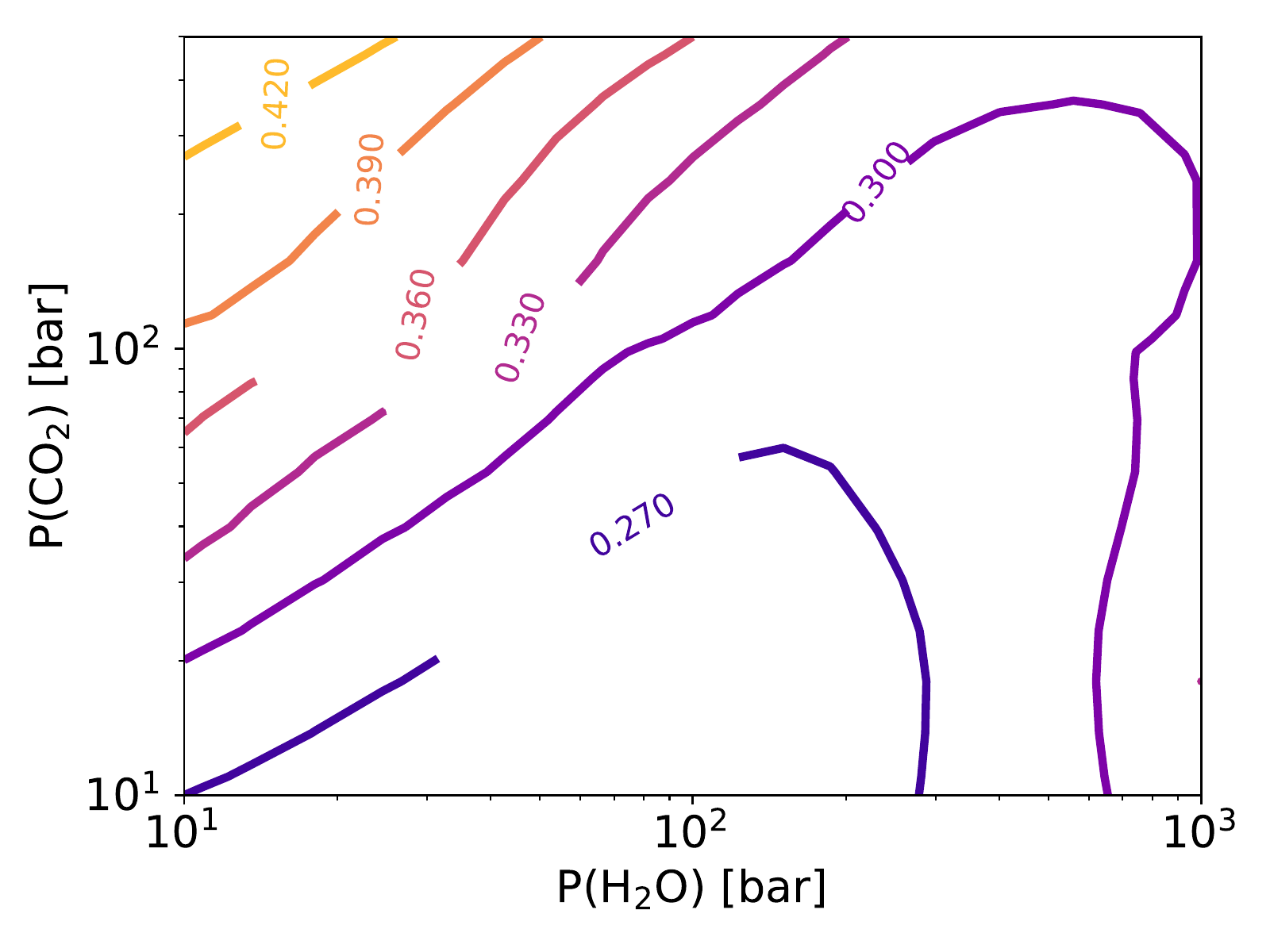}\hfill{}

\caption{$A_H$ contour plot with respect to
  $\mathrm{P}_{\mathrm{surf}}$(H$_2$O) and $\mathrm{P}_{\mathrm{surf}}$(CO$_2$),
  assuming a Sun-like star
  ($\mathrm{T}_*=5750\,\mathrm{K}$).\label{fig:Contour_3000K}}
\end{figure*}

For surface temperatures higher than the albedo transition
temperature, the albedo shows two distinct trends as we can see on
Fig.~\ref{fig:Contour_3000K}. Please remember that for
$T_{\mathrm{surf}} \gg T_A$, cloud scattering becomes negligible and
that only Rayleigh scattering contributes to the albedo. For very hot surface temperature, the cloud layer is indeed very thin when the atmosphere is water-dominates (Fig.~\ref{fig:Profil-de-temperature} left and Fig.~\ref{fig:humidity_profile} left) or even nonexistent for atmosphere CO$_2$-dominated (Fig.~\ref{fig:Profil-de-temperature} right and Fig.~\ref{fig:humidity_profile} right). When H$_2$O
partial pressure is relatively low, the albedo $A_H$ exhibits the same
behavior than $A_L$ in Fig.~\ref{fig:Contour_800K}. This behavior is
consistent with the work of \citet{KopparapuRamirezKastingEtAl2013}
who showed that the albedo increases from $0.3$ to $0.5$ with partial
pressure of CO$_2$ increasing from $1$ to $35\,\mathrm{bar}$. These
values are comparable to the albedos on Fig.~\ref{fig:Contour_3000K}
which range from $0.28$ to $0.40$. On the other hand, when H$_2$O
surface pressure is higher than about $10$ times than CO$_2$ surface
pressure, the albedo is increasing with $P_{\mathrm{surf}}$(H$_2$O)
instead, with little dependency with respect to
$P_{\mathrm{surf}}$(CO$_2$). This is because, for such high partial
pressures of H$_2$O, line opacity is quickly saturated, and most of
the reflectivity occurs in spectral windows of H$_2$O, so that water
opacity is mostly due to the H$_2$O-H$_2$O continuum. Also, the
effective Rayleigh scattering level (for $\tau_{\mathrm{ray}} \sim 1$)
occurs at a more or less constant pressure (close to a few bars), but
the temperature at these pressures is increasing with H$_2$O/CO$_2$
ratio due to the larger lapse rate of H$_2$O compared to CO$_2$
(keeping in mind that the temperature at vanishing pressures is always
close to $200\,\mathrm{K}$, see Fig.~\ref{fig:Profil-de-temperature}).
It appears that H$_2$O-H$_2$O continuum opacity is greatly reduced at
temperatures higher than about $500\,\mathrm{K}$, reached at these
pressures level in this H$_2$O/CO$_2$ domain, therefore yielding a
counter-intuitive increase in albedo with increasing
$P_{\mathrm{surf}}$(H$_2$O). This results are however considered not
very robust, since the details of H$_2$O-H$_2$O continuum opacity are
poorly known at such high temperatures (our tabulated values from MT\_CKD~3.0 are
capped at a temperature of $700\,\mathrm{K}$).

\subsection{Albedo vs. star temperature}

\begin{figure*}
\hfill{}\includegraphics[scale=0.8]{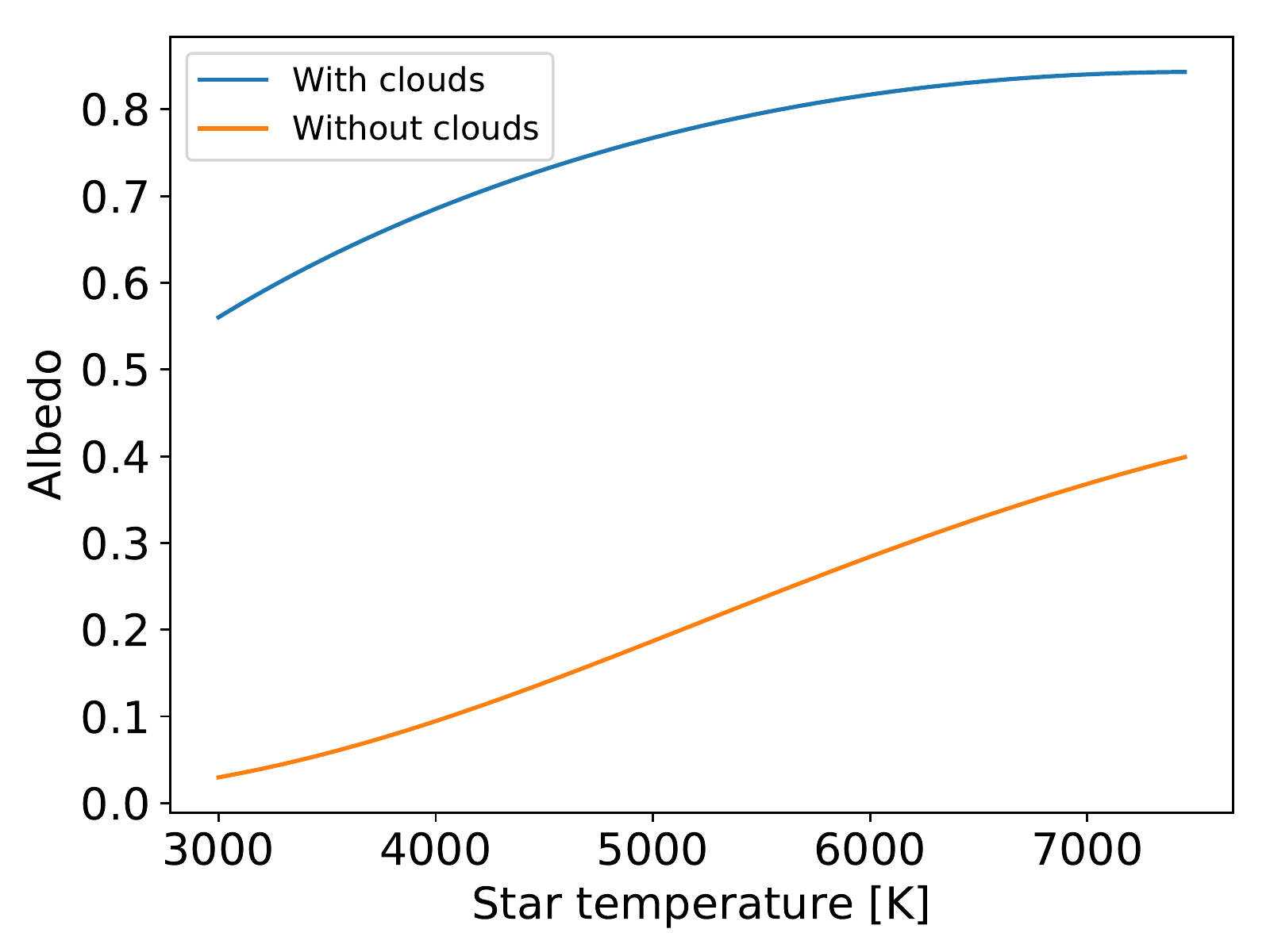}\hfill{}

\caption{Albedo at low surface temperature ($\mathrm{A}_\mathrm{L}$) with clouds (blue) and without clouds (orange) taken into account for a water-dominated atmosphere as a function of star temperature. Atmospheric inventory : $\mathrm{P}_\mathrm{surf}(\mathrm{H}_2\mathrm{O})=200\,\mathrm{bar};\,\mathrm{P}_\mathrm{surf}(\mathrm{CO}_2)=100\,\mathrm{bar}$. \label{fig:A_L_vs_Tstar}}
\end{figure*}

Since our modeled reflectivity is not constant in the visible-near IR
range, stellar temperature $\mathrm{T}_*$ obviously impacts planetary albedo.
We show in Fig.~\ref{fig:A_L_vs_Tstar} the Bond albedo of MO planets as a function of the star temperature for the standard atmosphere inventory ($\mathrm{P}_\mathrm{surf}(\mathrm{H}_2\mathrm{O})=200\,\mathrm{bar};\,\mathrm{P}_\mathrm{surf}(\mathrm{CO}_2)=100\,\mathrm{bar}$), with and without clouds, and for a surface temperature $\mathrm{T}_\mathrm{surf}=800\,\mathrm{K}$. As expected, the albedo is increasing with increasing stellar temperatures in both cases.  

The reasons for this behavior are twofold. Firstly, the
clouds absorb more and more at wavelengths longer than $2\,$\textmu m (see Tab.~\ref{tab:parametresnuages}). Secondly, Rayleigh scattering becomes negligible for wavelengths longer than about $1\,$\textmu m. Our results are moreover in agreement with
\citet{KopparapuRamirezKastingEtAl2013}, who previously showed that the Bond albedo increases with stellar temperature. We get Bond albedos of $0.03$ and $0.07$ for an atmosphere illuminated by $3000\,\mathrm{K}$ and $3800\,\mathrm{K}$
M-dwarfs respectively, and \citet{KopparapuRamirezKastingEtAl2013} find $0.01$ and $0.04$ for similar planets illuminated by $2600\,\mathrm{K}$ and $3800\,\mathrm{K}$ M-dwarfs, respectively. The small differences between their study and ours are probably due to differences in composition (their atmospheres contain almost no CO$_2$ unlike ours) and H$_2$O-H$_2$O continua values (MT\_CKD~3.0 in our study, BPS in
\citet{KopparapuRamirezKastingEtAl2013}).

According to Fig.~\ref{fig:A_L_vs_Tstar}, the increase of albedo in the cloudless case is comparatively stronger than the increase of the cloudy albedo. The latter also exhibits some saturation once the stellar effective temperature reaches about $6000\,\mathrm{K}$.
Indeed, our 1D model forces an altitude to the cloud layer according to the surface temperature of the planet. The Fig.~\ref{fig:A_L_vs_Tstar} shows a relatively cold planet ($\mathrm{T}_\mathrm{surf}=800\,\mathrm{K} \ll T_A$) which exhibits a thick and dense cloud layer located at low altitude. One should note that spectral reflectance of the clouds is considered equal to $1$ for wavelengths smaller than $2~$\textmu m (Tab.~\ref{tab:parametresnuages}). Hotter stars with emission peaks shifted to shorter wavelengths therefore results in an increased albedo. Once the star is hot enough, most of its emission falls below $2$~\textmu m. This explains the saturation of the albedo for hotter stars observed on Fig.~\ref{fig:A_L_vs_Tstar} when clouds are taken into account.
On the other hand, the albedo for cloudless atmosphere increases almost linearly with respect to the star temperature. Indeed, without clouds, only Rayleigh scattering contributes to the albedo increases, and Rayleigh scattering is more and more efficient at shorter wavelengths. Therefore, the impact of the star's temperature is significantly more dramatic for cloudless atmospheres.

\section{Limitations}
\label{sec:discus}


\subsection{Temperature profile}

In the current version of our model, the mesospheric temperature is fixed at $200\,\mathrm{K}$~\citep{Marcq2012,Marcqetal2017}. 
Our model does not take into account the heating induced by the visible and UV
radiation from the star, and therefore does not allow for the
emergence of a possible stratosphere (defined as an atmospheric layer
where $dT/dz > 0$). The temperature profile could affect the Bond albedo of the planet in an indirect way, through changes in the cloud opacity.

Note that the Bond albedo for cloudy atmosphere MO planets is significantly higher than in \citet{KopparapuRamirezKastingEtAl2013} (Figure 3b), because clouds are not taken into account in their work. It is tempting to believe that the Bond albedo calculated in our study could change the fate of planets going into runaway greenhouse, as such planets should behave in the same way as cool MO planets described in our work. As a planet warms up and goes into runaway greenhouse, a thick reflective cloud cover forms and stabilizes the climate of the planet. However, our calculations neglected 1) the effect of the incoming stellar radiation on the temperature profile and thus on cloud formation and 2) 3-D atmospheric circulation processes that can control the formation and properties of clouds \citep{Leconte2013}. Formation of clouds and their impact on the fate of planets going into runaway greenhouse could be tested in the future with 3-D Global Climate Model simulations taking into account clouds.

\subsection{Opacity data}

Although we regularly consider pressures over both critical pressures of H$_2$O and CO$_2$, we usually do so at temperatures much larger than their respective critical temperatures, so that the departure with ideal gas is limited, especially for CO$_2$. Also, for dense atmospheres most of backscattering occurs at relatively low pressures fully within our look-up tables, even more so when considering cloud opacity. It is nevertheless true that for relatively thin and hot atmospheres (therefore cloudless), we have non negligible contribution of layers hotter than 1000~K to the backscattering. In such a case our results are actually dependent of the poorly known opacities at these temperatures/pressures.

\subsection{Cloud parameterization}

Another limitation is the parameterization of clouds. We have
indeed considered current terrestrial type clouds following \citet{kasting1988}.
But these clouds are probably inappropriate for some temperature profiles, especially when the clouds are very high in the atmosphere. Plus, temperature near the top of the cloud layer is below 273~K, so that we
expect water ice clouds rather than liquid clouds. However, preliminary
testing with 10-micron ice particles instead of 10-micron liquid
water droplets only yield minor changes in resulting albedo (less than
2$\%$), with a slightly higher albedo around G-stars and slightly lower
around M-stars, consistent with the lower single scattering albedo of
water ice compared to liquid water in the near infrared. We should also
note that altering cloud microphysics (vertical mass loading profile,
size distribution, contaminants, etc.) could have a more significant
impact on our results. This highlights the need for a self-consistent
microphysical scheme in future versions of our model.

\subsection{Mesopheric temperature}

An important point concerns the mesospheric temperature. As stated in Section~2, the mesospheric temperature is fixed in our model at 200~K. In fact, depending on the temperature of the host star and the irradiation received by the planet, the mesosphere could be more or less heated. We tested our results for three different mesospheric temperatures (150~K, 200~K and 250~K) and found that the bond albedo is the same for the cooler mesospheric temperature (150~K). For a mesospheric temperature of 250~K, the albedo decreases by less than 2-3\% , due to an increase in mesospheric humidity.

\subsection{Realistic star spectra}
\label{sssec:star_spec}

To calculate the MO planetary albedo, we made the approximation that the stars emitted like blackbodies. 

However, while this assumption is quite correct for G-stars, it is much less so for colder stars (M-dwarfs) because the atmospheres of these stars are more
complex. For example, Proxima Centauri's spectrum, which is now well
constrained by observations, displays an absorber in its atmosphere,
especially in the visible around $0.4$ and $0.6\,$\textmu m and in the
near UV around $0.3\,$\textmu m~\citep{Ribas2017}.
Therefore, the incident spectrum flux at the top of the atmosphere of
the MO planet contains less visible and UV flux compared to a black
body spectrum which should further lower the albedo of the planet in
this case. 

\section{Conclusion}

This paper presents a simple investigation of the albedo of MO
planets. Our updated model allowed us to calculate the albedo of MO
planets for surface temperatures between $800$ and $3000\,\mathrm{K}$
for a large array of H$_2$O-CO$_2$ atmospheric inventories and for various
star spectral type. This study has evidenced the strong influence of
clouds (whenever present) on the Bond albedo. The albedo can become very high (larger than $0.9$) when the clouds are optically thick. We have also shown that the composition of
the atmosphere, and in particular the mixing ratio CO$_2$/H$_2$O,
plays a key role on the albedo. Our work is in agreement with those of
\citet{KopparapuRamirezKastingEtAl2013} and
\citet{Hamano2015}. Besides, we have shown that the spectral type of
the host star strongly influences the resulting albedo in the same way
that \citet{KopparapuRamirezKastingEtAl2013} did.  Our results finally
show that the variations of the albedo with respect to surface
temperature can be parameterized with a precision better than $5\%$
using a small number of quantities (asymptotic values at high and low
surface temperature, transition temperature).

The next step of this research project is to make all the 
improvements presented above and then incorporate them all into a Global Climate Model (GCM) to simulate the cooling phase of young telluric planets. We could thus be able to better simulate the transition 
phase when the internal heat flux of the planet becomes negligible in 
front of the received stellar flux and thus we could have a more precise 
idea of the potential habitability of these planets. A second major challenge 
would be to use the spectral data and the albedo provided by our 
atmospheric model to compare them with the instrumental data expected 
from the space missions that will very soon study exoplanets such as the 
James-Webb Space Telescope (JWST) which will be launched by NASA/ESA in 2020.

\section{Acknowledgement}

This project has received funding from the European Research Council (ERC)
under the European Union’s Horizon 2020 research and innovation programme
(grant agreement n$^\circ$ 679030/WHIPLASH). EM acknowledges the support of INSU/PNP.

\bibliographystyle{authordate1}
\bibliography{biblio_jabref_stage_M2}

\end{document}